\documentclass[pre,twocolumn,aps,floats,showpacs,superscriptaddress]{revtex4}

\usepackage{amsmath}
\usepackage{amssymb}
\usepackage{latexsym}
\usepackage{bm}
   
\usepackage{graphicx}
\usepackage{epsfig}
\usepackage{epic}
\usepackage{eepic}

\newcommand{\xx}{{\bf x}}
\newcommand{\vv}{{\bf v}}

\newcommand{\bb}{\textbf}
\newcommand{\avk}{\langle k \rangle}

\begin{document}
\bibliographystyle{prsty}
\author{Cecilia Nardini} 
\affiliation{LPT, CNRS, UMR 8627, and 
Univ Paris-Sud, Orsay, F-91405  (France)}
\affiliation{Universit\'a di Padova, dipartimento di Fisica "G.
Galilei" (Italy)}
\author{Bal\'azs Kozma} 
\affiliation{LPT, CNRS, UMR 8627, and 
Univ Paris-Sud, Orsay, F-91405  (France)}
\author{Alain Barrat} 
\affiliation{LPT, CNRS, UMR 8627, and 
Univ Paris-Sud, Orsay, F-91405  (France)}
\affiliation{Complex Networks Lagrange Laboratory, ISI Foundation, 
Turin, Italy}

\title{Who's talking first? Consensus or lack thereof in coevolving
opinion formation models} \widetext

\begin{abstract}
  We investigate different opinion formation models on adaptive network
  topologies. Depending on the dynamical process, rewiring can
  either (i) lead to the elimination of interactions between agents in
  different states, and accelerate the convergence to a consensus state or
  break the network in non-interacting groups or (ii) counter-intuitively,
  favor the existence of diverse interacting groups for exponentially long
  times.  The mean-field analysis allows to elucidate the mechanisms at play.
  Strikingly, allowing the interacting agents to bear more than one opinion at
  the same time drastically changes the model's behavior and leads to fast
  consensus.
\end{abstract}
\date{\today}
\pacs{89.75.-k, -87.23.Ge, 05.40.-a}


\maketitle

In the recent years, agent based models have been more and more used
in the area of social sciences. Through a rather simple
modeling approach for the individual processes of social influence,
these models focus on the emergence of social behavior at the global population
level. Statistical physics models and tools provide therefore a
natural framework for such studies, and have been widely applied,
leading to the appearance of the field called
sociophysics (see \cite{review_CFL} for a recent review on the
application of statistical physics models to social dynamics).

The growing field of complex networks
\cite{Dorogovtsev:2003a,Newman:2003b,pastorsatorras2004} has moreover
allowed to obtain a better knowledge of social
networks~\cite{Granovetter:1973,Wasserman:1994}, and in particular to
show that the typical topology of the networks on which social agents
interact is not regular. Many studies have therefore considered the
evolution of models of interacting agents when agents are embedded on
more realistic networks, and studied the influence of various complex
topologies on the corresponding dynamical behavior
\cite{Boccaletti:2006}. An additional feature of networks, that may
have a strong impact on the model's behavior, lies in their dynamical
nature. They may indeed evolve on various timescales, and the
evolution of the topology and the dynamical processes can drive each
other with complex feedback effects. Studies of this coevolution
are more recent and still limited
\cite{Zimmermann:2004,Gross:2006,Ehrhardt:2006,Gil:2006,Centola:2006,Holme:2006,Garlaschelli:2007,Kozma:2007,Benczik:2007,Vazquez:2007}, with many open issues.

In this Letter, we provide new insights into such feedback effects by an
investigation of Voter-like models (VM),
in which agents update their opinions by imitating their neighbors,
and can also break and establish connections with other agents.  More
precisely, we show how slight modifications in the evolution rule,
which have minor consequences if the topology of interactions is
kept fixed, can change drastically the model's behavior as
soon as the topology can evolve on the same timescale as the agents'
opinions. On the other hand, the simple fact of allowing agents to
have several opinions at the same time, in the spirit of the Naming
Game \cite{baronka} or of the AB model \cite{Castello:2006}, leads to
more robust behavior.

The Voter model \cite{Krapivsky:1992} considers a population of size
$N$ in which each individual $i$ has an opinion characterized by a
binary variable $s_i=\pm 1$: only two opposite opinions are here
allowed (for example a political choice between two parties)
\footnote{We have also investigated the case in which a finite number
of opinions is allowed, with the same results.}.  
Starting from a random configuration of
opinions, the dynamical evolution of the {\em direct} VM (d-VM) is the
following: at each elementary step, an agent ($i$) is randomly
selected, chooses one of its neighbors ($j$) at random and adopts its
opinion, i.e. $s_i$ is set equal to $s_j$ (one timestep consists
of $N$ such updates). In the {\em reverse} case
(r-VM), the first agent $i$ instead convinces its neighbor $j$ ($s_j$
is set equal to $s_i$). The distinction between d- and r-VM
is necessary since the two interacting nodes do not play the
same role. Moreover, the degrees of the first and the second chosen
nodes have different distributions, and the second is a large-degree
node with larger probability \cite{pastorsatorras2004}. The asymmetry
in the opinion update between the two interacting nodes can then
couple to the asymmetry between a randomly-chosen node and its
randomly-chosen neighbor, leading to different dynamical
properties. No important difference is expected on homogeneous
networks but, on heterogeneous networks, the probability for a hub to
update its state will vary strongly from one rule to the other. The
basic imitation process of the VM mimics the homogenization of
opinions but, since interactions are binary and random, do not
guarantee the convergence to a uniform state. Since a
consensus in which all individuals share the same opinion is an
absorbing state of the dynamics, any finite population reaches a
consensus, but the time needed $t_c(N)$ depends on its size $N$ and on
the topology of interactions, and diverges as $N \to \infty$.  
On static networks, $t_c(N)$ grows as a power-law of $N$, with an
exponent depending on the degree distribution, and on the updating rule
\cite{Sood:2005,Suchecki:2005,Castellano:2005}. On homogeneous
networks in particular, $t_c(N) \propto N$ for both d- and r-VM.

In this Letter, we consider the scenario in which agents can rewire their
``unsatisfied'' connections. More precisely, the initial configuration is
given by a random homogeneous network of interacting agents, with average
number of neighbors $\avk$ and random opinions.  At each timestep, an agent
$i$ and one of its neighbors $j$ are chosen. With probability $\Phi$, an
attempt to rewire the link is made, if $s_i \ne s_j$. A new agent $k$ is then
chosen at random and the link $(i,j)$ is rewired to $(i,k)$\footnote{Another
  possibility is that the rewired link becomes $(j,k)$; the same phenomenology
  is then observed.}. With probability $1-\Phi$, an opinion update takes place
instead. The rewiring, which conserves the total number of links, is made at
random: the new link is established without prior knowledge of the new
neighbor's opinion.

If the frequency of rewirings is small ($\Phi \to 0$), the system
still reaches a global consensus and the network remains
connected. For fast rewiring rates on the other hand, ($\Phi \to 1$)
the network breaks into (typically two for the VM) separate connected
components, each one with a local consensus. These two regimes are
separated by a non-equilibrium phase transition at a critical value
$\Phi_c$ of the rewiring probability. Similar transitions have already
been reported in co-evolving models of opinion formation
\cite{Holme:2006,Kozma:2007,Vazquez:2007} and we will not focus on
this aspect here.

\begin{figure}[tb]
\centering
\rotatebox{0}{
\includegraphics[width=.4\textwidth]{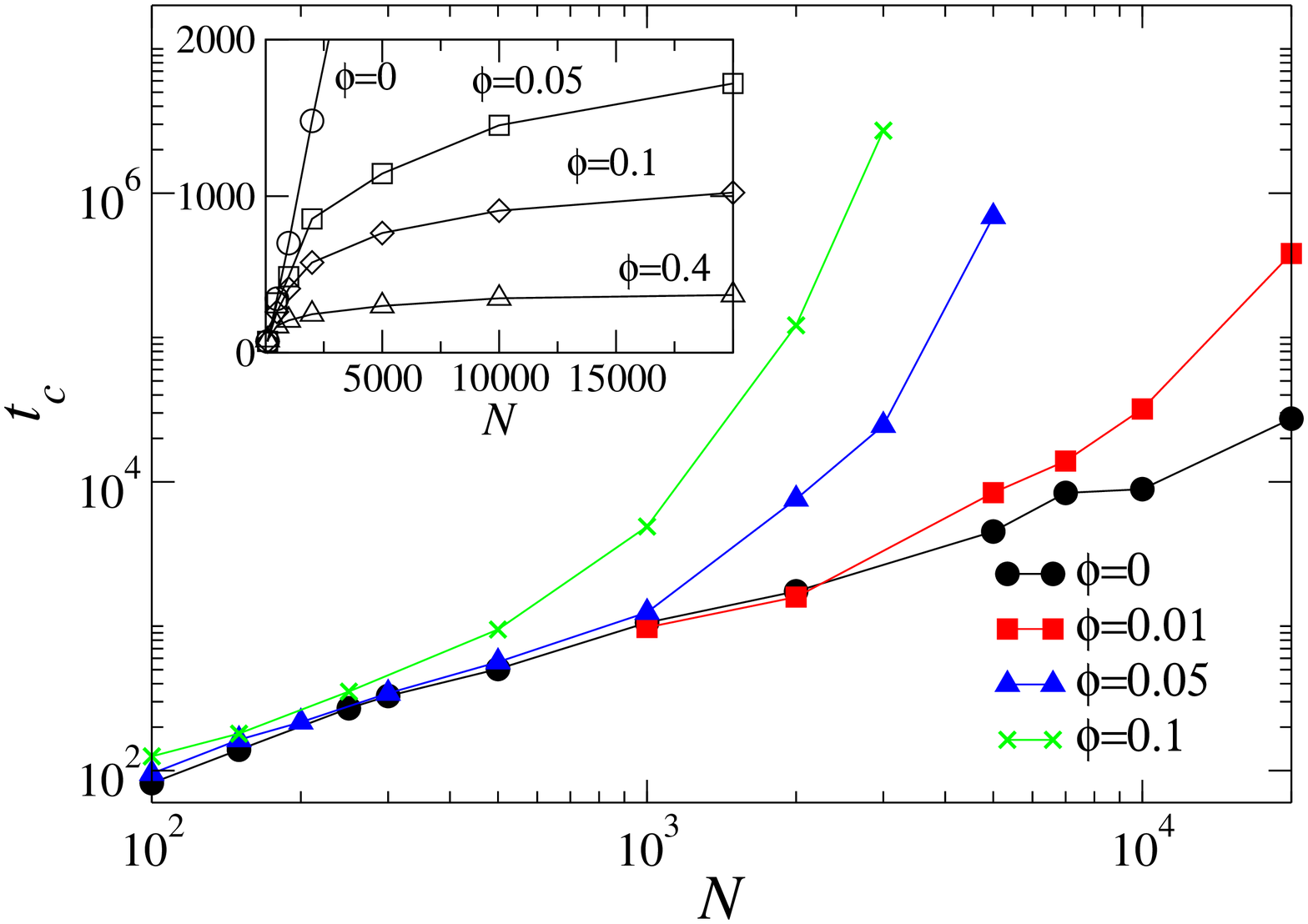}
}
\vspace*{-0.0cm}
\caption{(Color online)
Convergence time for the r-VM as a function of the population size,
for various rewiring probabilities. Throughout this paper, 
the data from simulations 
were averaged over $100$ realizations of the system.
Inset: same for the d-VM. Note the difference of scales.
\label{fig:tc_VM}
}
\end{figure}

A more surprising aspect of the dynamics is revealed by the behavior of the
convergence time $t_c(N)$, which grows linearly with $N$ on a static network
for both the d- and the r-VM \cite{Sood:2005,Castellano:2005}. Strikingly, the
network's adaptivity has completely opposed effects in these models
(Fig. \ref{fig:tc_VM}).  Consensus is strongly favored in the d-VM, for which
$t_c(N)$ becomes $\propto \ln N$ \footnote{Close to the transition $\Phi \sim
  \Phi_c$, $t_c(N) \sim N^a$, $a \approx 0.43$.}; in contrast, for the r-VM
$t_c(N)$ grows exponentially with the system size. The system therefore
remains for exponentially long times in a state in which two groups of
different opinions co-exist and remain connected to each other. It is
noteworthy that the system therefore is not frozen, with agents continuously
updating their links and opinions.

\begin{figure}[thb]
\centering
\rotatebox{0}{
\includegraphics[width=.4\textwidth]{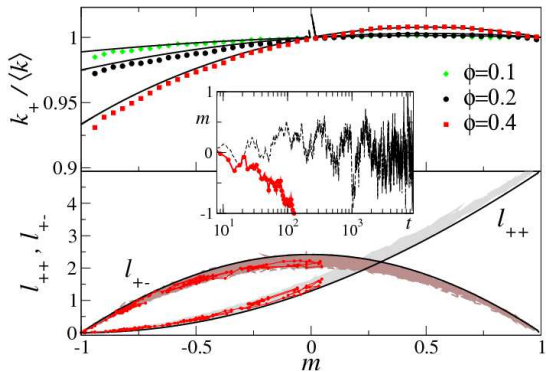}
}
\vspace*{-0.0cm}
\caption{(Color online) Top: $k_+/\avk$ vs $m$ for the VM. 
The symbols correspond to averages obtained from numerical simulations
with $N=1000$, $\avk=10$, while the continuous lines 
are the numerical solution of the MF equations for the evolution of
$\xx=(m,l_{+-},l_{++})$, starting from initial conditions with
$m$ close to $0$. Bottom: 
$l_{++}$ and $l_{+-}$ vs $m$. 
The continuous black lines correspond to the numerical
solution of the MF equations. The red symbols and the
grey and brown lines
correspond to single runs
of the d- and r-VM, respectively ($N=500$, $\avk=10$, $\Phi=0.4$). 
The inset in the middle shows the evolution of $m$ for the same runs 
(red symbols for the d-VM and dashed line for the r-VM). 
\label{fig:k_VM}
}
\end{figure}

In order to understand the different behavior of the d- and r-VM on
adaptive networks, we note that the state of the system is characterized by
three independent quantities: (i) the density $n_+$ of agents with
opinion $+1$, or equivalently the magnetization $m=n_+ - n_-$ ($n_-=1-n_+$ is
the density of agents with opinion $-1$); (ii) the number of links joining
agents in the $+$ opinion, $N l_{++}$; and (iii) the number of links joining
agents of opposite opinions, i.e. of {\em active} links $N l_{-+}=N l_{+-}$
(since the total number of links is preserved, $\avk /2 = l_{++} + l_{+-} +
l_{--}$). At the mean-field (MF) level, we can derive the evolution equation
of these quantities. Let us first consider the magnetization: it changes of
$-2/N$ when an agent changes its states from $+$ to $-$, and of $+2/N$ in the
opposite case. For the d-VM, the probability of the first event is
proportional to the density $n_+$ of agents in the $+$ state, times the
probability that it chooses to interact with a neighbor that has $-$
opinion, i.e. $k_{+-}/k_+$ where $k_+$ is the average degree of a $+$ node,
and $k_{+-}=l_{+-}/n_+$ is the average number of $-$ neighbors of a $+$ node.
The probability of the second event ($- \to +$) is obtained in the same way,
and finally
 \begin{equation}
 \left< \frac{d m}{d t} \right>_{\mbox{\scriptsize{dVM}}} 
= - \frac{2(1-\Phi)}{N}  \ l_{+-} \left( \frac{1}{k_{+}} - \frac{1}{k_{-}}
\right) \ .
\label{eq:VM_dmdt}
\end{equation}
In the case of the r-VM, the probabilities of the two processes are
simply interchanged: $\left< d m/d t
\right>_{\mbox{\scriptsize{rVM}}}= - \left< d m/d t
\right>_{\mbox{\scriptsize{dVM}}}$. 
On an adaptive network, it is essential to distinguish $k_+$ from $k_-$: as
shown in Fig. \ref{fig:k_VM}, one has indeed $k_+ > \avk > k_-$ if $n_+ >
n_-$. In other words, the nodes of the majority opinion have more neighbors.
This is a simple consequence of the rewiring dynamics: if $m > 0$, any
rewiring event $(i,j) \to (i,k)$ has a higher chance to randomly pick a $+$
node as a new neighbor due to their larger number. Therefore, nodes of the
larger group gain new links with larger probability. Equation
(\ref{eq:VM_dmdt}) then immediately shows that for $m>0$, $\left< d m/d t
\right>_{\mbox{dVM}} > 0$ and $\left< d m/d t \right>_{\mbox{rVM}} < 0$.  In
summary, the coevolution of opinions and topology generates a positive
feedback for the d-VM driving the system to a consensus state, $m_{stable}=\pm
1$, and a negative feedback for the r-VM resulting in $m_{stable}=0$. This
readily explains the strong differences between these models. For the d-VM the
adaptivity leads to an accelerated consensus, while it hinders the convergence
for the r-VM and keeps the system in a dynamically evolving state with zero
average magnetization.

It is moreover possible to write the evolution equations for the various types
of links. It is easy to understand that, according to the model's definition,
the vector $\xx=(m,l_{+-},l_{++})$ can evolve in 4 ways at each elementary
update: $\xx \to \xx+\vv^a$, $a=1,\cdots,4$, with respective probabilities
$w^a$. Let us start with the d-VM.  The displacement vectors and the
associated probabilities read then: $N \vv^1 = \bb{(} 2 \bb{,} \ k_{--}-k_{-+}
\bb{,} \ k_{-+} \bb{)}$, $w_1 = (1-\Phi) \ n_- \ {k_{-+}}/{k_-}$; $N \vv^2 =
\bb{(}-2 \bb{,} \ k_{++}-k_{+-}\bb{,} \ -k_{++} \bb{)}$, $w_2 = (1-\Phi) \ n_+
{k_{+-}}/{k_+}$; $N \vv^3 = \bb{(}0 \bb{,} \ -1 \bb{,} \ 0 \bb{)}$, $w_3 =
\Phi \ n_-^2 \ {k_{-+}}/{k_-}$; $N \vv^4 = \bb{(}0 \bb{,} \ -1 \bb{,} \ +1
\bb{)}$, $w_4 = \Phi \ n_+^2 \ {k_{+-}}/{k_+}$.  $\vv^1$ and $\vv^2$
correspond to opinion changes, for which the change in magnetization ($\pm
2/N$) is associated with changes in the densities of links. For example, when
a $-$ node is transformed to $+$, its $--$ links become $+-$ and its $+-$
links become $++$ ones (hence $l_{+-}$ varies of $(k_{--}-k_{+-})/N$). The
corresponding probabilities $w_1$ and $w_2$ are obtained as for
Eq.~(\ref{eq:VM_dmdt}). $\vv^3$ and $\vv^4$ correspond to rewiring events:
when a $+-$ link is rewired it can be either transformed to $--$ ($\vv^3$) or
to $++$ ($\vv^4$).  For the r-VM, the displacement vectors are exactly the
same as for the d-VM, but the transition probabilities $w_1$ and $w_2$ are
interchanged. $w_3$ and $w_4$ remain the same since the rewiring rules 
are the same for both models. Figure \ref{fig:k_VM} shows the result of the
numerical integration of the evolution equations $d\xx/dt=\sum_a \vv^a w^a$,
compared with numerical simulations of the models. It is clear that these
equations correctly account for the difference between $k_+$ and $k_-$ and for
the system's evolution in the phase space. Of course, the real systems are
moreover submitted to fluctuations that are not taken into account in the MF
description. In particular, looking at single runs (Fig.
\ref{fig:k_VM}) shows clearly the difference between the d- and r-VM. For the
d-VM, the density of active links decreases rapidly to $0$ and the
system is driven to one of the consensus states. For the r-VM on the
contrary, the system performs a random walk in a sort of 
potential well around $m=0$ with a non-zero density of active links,
which ends only because of a finite-size fluctuation which leads
it into one of the absorbing boundaries at $m=\pm 1$.

\begin{figure}[thb]
\centering
\rotatebox{0}{
\includegraphics[width=.4\textwidth]{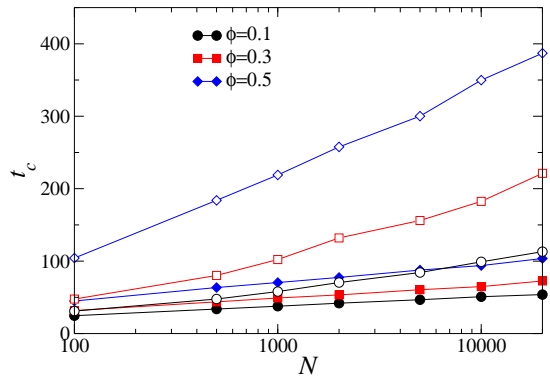}
}
\vspace*{-0.0cm}
\caption{Convergence time for the direct (filled symbols) and
reverse (open symbols) NG.
\label{fig:tc_NG}
}
\end{figure}

\begin{figure}[tb]
\centering
\rotatebox{0}{
\includegraphics[width=.4\textwidth]{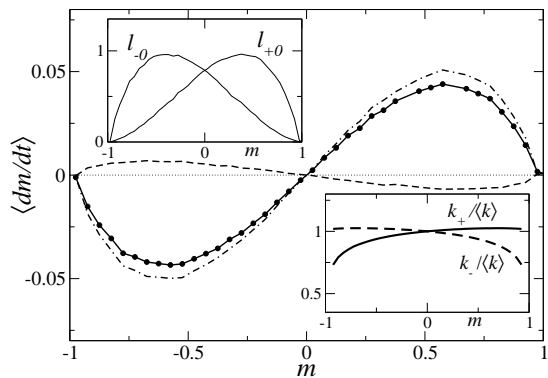}
}
\vspace*{-0.0cm}
\caption{$\langle dm/dt \rangle$ vs $m$ (symbols) 
for the r-NG. According to Eq. 
(\ref{eq:rNG_dmdt}), changes in the magnetization come from r-VM-like
interactions (dashed line) and those mediated by the $0$-links, $l_{+0}$ and
$l_{-0}$ (dash-dotted line).
The upper inset gives $l_{+0}$ and $l_{-0}$, the other 
$k_+/\avk$, $k_-/\avk$. $\Phi=0.2$, $\avk=10$, $N=10^4$.
\label{fig:NG_dmdt}
}
\end{figure}

Let us now consider that agents cannot pass directly from one opinion
to another, but can keep both opinions in their ``memory'', being then
in an intermediate state that we call $0$. This is the case in the
Naming Game (NG) model, in which agents try to agree on the name to
assign to a given object \cite{baronka}, or also of the AB model
\cite{Castello:2006}. If only two names are available (that we can
call $+$ and $-$ for simplicity), the dynamical rules of the {\em
direct} NG (d-NG) are the following: at each timestep, an agent $i$ and one
of its neighbors, $j$, are chosen at random to be respectively Hearer
(H) and Speaker (S). S proposes a name to H. If S has both names in
memory, it chooses one at random. Let us suppose for instance that S
proposes $+$. If H does not know the name uttered (i.e., it is in
state $-$), it absorbs this possibility by changing to the
intermediate state, $0$. If H instead has the name in memory (i.e., it
is in state $+$ or $0$), the interaction is successful and both H and S
agree on this particular name and set in state $+$ after the
interaction. In the reverse case (r-NG), the first randomly selected
agent is S and its neighbor is H, and the update rules remain the
same \footnote{In the AB model, only H changes its state; we have
checked that this modification has no influence on the overall
behavior of the system described below.}. When agents interact on a
static topology, these dynamical rules lead to a global consensus. On
homogeneous networks, we obtain $t_c(N) \sim \ln N$ (while $t_c(N)
\sim N$ for the VM). The difference between the two models is due to
the fact that in the VM, consensus is reached by a finite-size
fluctuation of the average magnetization while in the NG, consensus is
reached due to the surface-tension introduced by the $0$ states, which
tends to minimize the interface between the agents of different
opinions and hence drive the system to a homogeneous consensus state
\cite{Baronka:2006,Castello:2006}. For adaptive networks, 
Fig. \ref{fig:tc_NG} clearly shows that the convergence
time remains logarithmic for both the direct and reverse version, even if
the r-NG is slower. 
The MF analysis allows to understand this strong difference
with the VM. We can indeed write the evolution equation for
the magnetization $n_+ - n_-$, by introducing the average degree of
$0$ nodes $k_0$ and the density of $+0$ and $-0$ links, as
\begin{eqnarray}
 \left< \frac{d m}{d t} \right>_{\mbox{\scriptsize{dNG}}} 
&=& \frac12 \left< \frac{d m}{d t} \right>_{\mbox{\scriptsize{dVM}}} 
+ \frac{1-\Phi}{N} \left( \frac{l_{+0}}{k_0} - \frac{l_{-0}}{k_0} \right)
\label{eq:dNG_dmdt} \\
 \left< \frac{d m}{d t} \right>_{\mbox{\scriptsize{rNG}}} &=& \frac12
\left< \frac{d m}{d t} \right>_{\mbox{\scriptsize{rVM}}} + \frac{1-\Phi}{N}\left(
\frac{l_{+0}}{k_+} - \frac{l_{-0}}{k_-} \right) \ .
\label{eq:rNG_dmdt}
\end{eqnarray}
The first terms on the rhs represent the change in the magnetization
mediated by the $l_{+-}$ links. The factor $1/2$ stems from the fact
that $+$ and $-$ nodes are not transformed instantly to their opposite
counterpart but to the intermediate state $0$. The remaining terms
correspond to the transformation of the $0$ nodes to $\pm$ ones. For
example, in the d-NG, the second term on the right-hand side is
generated by the process when a $0$ node is converted to $+$ by first
picking a $0$ node, with probability $n_0$, then one of its
$+$ neighbors, with probability $k_{0+}/k_0$, and so on. Even though
the first terms in Eq.-s~(\ref{eq:dNG_dmdt}) and (\ref{eq:rNG_dmdt})
change sign for the d- and the r- variants of the NG just as
for the VM, this effect is suppressed by the remaining terms
associated with the transitions $(0 \to \pm)$ which will always
generate a positive feedback to the change of magnetization. As shown
in Fig. \ref{fig:NG_dmdt} indeed, $l_{+0}-l_{-0}$ is of the sign of
$m$, which is expected since then $n_+ > n_-$.  This effect overcomes
the difference between $k_+$ and $k_-$; as a result, $\langle d m/dt
\rangle$ remains of $m$'s sign even in the r-NG, leading to
logarithmic convergence times.  Very interestingly, the possibility
for agents to remain in an intermediate state before updating their
opinion strongly enhances the trend towards consensus.

In summary, we have shown how slight modifications of the interaction
rules can have drastic consequences in the global behavior of opinion
formation models in the case of dynamically evolving networks. In the
case of the paradigmatic Voter model, adaptivity of the network can
either accelerate the convergence to consensus, or on the contrary
hinders it strongly, by maintaining the system in a dynamically
evolving state for exponentially long times. A mean-field analysis
allows to account for such differences, which are due to the coupling
of the asymmetry between the interacting agents to the asymmetry in
their degrees. Such coupling is known to change the scaling of the
convergence time in heterogeneous static networks, which however
remains a power-law of time \cite{Sood:2005,Castellano:2005}, and in
fact does not have consequences in homogeneous networks.  In strong
contrast, and even if the adaptive network remains homogeneous, the
fact that the majority has a slightly larger average degree suffices
to change from a very fast convergence in logarithmic 
time for the d-VM to a dynamical state surviving for exponentially long times
for the r-VM. Interestingly, if the agents cannot change opinion so
easily, and have to go through an intermediate state, such as in the
NG or AB models, convergence to consensus is enhanced also for
adaptive networks, and irrespective of the order of interactions
(d-NG vs. r-NG). The connections with nodes in the
intermediate state determine then the dominant evolution of the
magnetization, leading to a more robust behavior.

{\em Acknowledgements ---} BK and AB
are partially supported by the EU
under contract 001907 (DELIS).

\end{document}